\documentclass[sigconf]{acmart}
\usepackage{hyperref}

\hypersetup{
    colorlinks=true,
    allcolors=blue
}
\setcopyright{none}
\renewcommand\footnotetextcopyrightpermission[1]{}

\fancypagestyle{standardpagestyle}{
  \fancyhf{} % clear everything
  \fancyhead[C]{\textit{My Paper Title}} % center header
  \fancyfoot[C]{\thepage} % page number (optional)
  
}
\fancypagestyle{firstpagestyle}{
  \fancyhf{}
  \fancyhead[C]{\textit{My Paper Title}} % or leave blank for no first-page header
  \fancyfoot[C]{\thepage}
  
}

\usepackage{booktabs}
\usepackage{tabularx}
\usepackage{stfloats}
\begin{document}

\title{Beyond Citations: A Cross-Domain Metric for Dataset Impact and Shareability}

\author{Smitha Muthya Sudheendra}
\authornote{Authors contributed equally}
\affiliation{%
  \institution{University of Minnesota,}
  \city{Twin Cities,}
  \state{Minnesota}
  \country{USA}
}
\email{muthy009@umn.edu}

\author{Zhongxing Zhang}
\authornotemark[1]
\affiliation{%
  \institution{University of Minnesota}
  \city{Twin Cities}
  \state{Minnesota}
  \country{USA}
}
\email{zhan8889@umn.edu }

\author{Wenwen Cao}
\authornotemark[1]
\affiliation{%
  \institution{University of Minnesota}
  \city{Twin Cities}
  \state{Minnesota}
  \country{USA}
}
\email{cao00428@umn.edu}

\author{Jisu Huh}
\affiliation{%
  \institution{University of Minnesota}
  \city{Twin Cities}
  \state{Minnesota}
  \country{USA}
}
\email{jhuh@umn.edu}

\author{Jaideep Srivastava}
\affiliation{%
  \institution{University of Minnesota}
  \city{Twin Cities}
  \state{Minnesota}
  \country{USA}
}
\email{srivasta@umn.edu}

\begin{abstract}
  The scientific community increasingly relies on open data sharing, yet existing metrics inadequately capture the true impact of datasets as research outputs. Traditional measures, such as the h-index, focus on publications and citations but fail to account for dataset accessibility, reuse, and cross-disciplinary influence. We propose the X-index, a novel author-level metric that quantifies the value of data contributions through a two-step process: (i) computing a dataset-level value score (V-score) that integrates breadth of reuse, FAIRness, citation impact, and transitive reuse depth, and (ii) aggregating V-scores into an author-level X-index. Using datasets from computational social science, medicine, and crisis communication, we validate our approach against expert ratings, achieving a strong correlation. Our results demonstrate that the X-index provides a transparent, scalable, and low-cost framework for assessing data-sharing practices and incentivizing open science. The X-index encourages sustainable data-sharing practices and gives institutions, funders, and platforms a tangible way to acknowledge the lasting influence of research datasets.
\end{abstract}


\maketitle

\section{Introduction}
The growing importance of open data in science and medicine highlights the need for evaluation systems that properly recognize dataset contributions. While traditional metrics such as the \emph{H-index} \cite{hirsch2005index} capture scholarly productivity through publications and citations, they overlook the impact of datasets, their reuse across domains, and the collaborative nature of data sharing. Despite these advances, the systems used to reward researchers remain primarily anchored to publication-based metrics, such as the h-index, that emphasize articles and their citations while ignoring the broader value and influence of datasets.

While open science thrives on data availability and reuse, researchers receive little professional recognition for sharing datasets. As a result, incentive systems fail to encourage sustainable data stewardship, interdisciplinary reuse, or dataset quality improvements guided by FAIR principles (Findable, Accessible, Interoperable, Reusable).

To address this gap, we propose the \emph{X-index}, a novel metric designed to quantify the overall value of an author’s dataset contributions. At the core of the framework is the dataset-level \emph{value score} (V-score), which integrates four dimensions: dataset breadth (reuse across domains), quality (measured by FAIR principles), citation impact, and reuse depth (decay-weighted transitive citations). These V-scores are then aggregated into an author-level X-index. A preliminary evaluation shows that the X-index strongly correlates with expert assessments of dataset impact.

Our contributions are threefold:
\begin{itemize}
    \item We introduce the X-index, a formalized metric for recognizing dataset contributions beyond traditional publication-based measures.
    \item We integrate four critical dimensions, breadth, quality, impact of citations, and depth of reuse, into a unified and transparent evaluation framework.
    \item We provide initial validation showing a strong correlation between X-index scores and expert judgments, demonstrating feasibility for large-scale adoption.
\end{itemize}

\section{Background}
Research impact has long been measured using publication-based bibliometric metrics, most notably the h-index  \cite{hirsch2005index}, which captures productivity and citation counts and is widely used across disciplines. The h-index has become one of the most widely used bibliometric indices for evaluating a researcher's productivity and impact. It is defined as the number h such that a scholar has published h papers, each of which has been cited at least h times. The h-index attempts to balance both the quantity and influence of scientific output in a single, intuitive metric. Despite its popularity, the h-index has been widely criticized for several limitations. It disregards citations beyond the h-core, meaning that highly cited papers do not receive additional weight once they contribute to the h count. It also fails to account for author contributions in multi-authored works, the age of publications or citations, and discipline-specific citation practices. Jin et al.~\cite{jin2006h} proposed the A-index, which is defined as the average number of citations among papers within the h-core. This was later extended through the R-index \cite{jin2007r}, which computes the square root of the total citations in the h-core. An additional variant, the $R_m$-index \cite{panaretos2009assessing}, uses the square roots of individual citation counts to reduce the impact of extreme citation outliers. Though these indices focus on performance by reflecting on citation, these indices ignore the "tail" of less cited papers. Additionally, they may amplify the influence of few outlier publications. 

The g-index\cite{egghe2006theory}, extends the h-index by giving more weight to highly cited papers. It is defined as the largest number g such that the top g papers have together received at least $g^2$ citations. The primary limitation of the g-index is that it can be disproportionately influenced by a few highly cited papers, potentially overstating a researcher's overall impact. Additionally, like the h-index, it does not account for co-authorship, publication age, or disciplinary citation norms. Time-adjusted indices attempt to account for career length and publication age. The m-quotient \cite{hirsch2005index} normalizes the h-index by the number of years since the researcher’s first publication. The AR-index \cite{jin2007r} further adjusts for the age of each paper in the h-core, thereby emphasizing recent productivity. The main limitation of the time-adjusted indices is that the selection of a starting year can be arbitrary, and the index may undervalue long-term impact or seminal early work that continues to be cited.

The OmicsDI index \cite{perez2019omicsdi} evaluates dataset impact by aggregating metrics such as views, downloads, citations, and cross-dataset reuse. By incorporating multiple dimensions of dataset interaction, it emphasizes both visibility and reusability across repositories. However, its reliance on repository-level metrics can introduce bias based on platform visibility, and it may underrepresent impact for datasets used outside tracked infrastructures or reused in informal settings. The U-index \cite{callahan2018u} introduces a distinction between usage citations—instances where a resource is actively employed and awareness citations, where a resource is merely mentioned. This separation allows for more credit assignment and discourages citation inflation through passive referencing. Nonetheless, the U-index requires detailed citation context parsing, which may be error-prone or inconsistent across articles, and it depends heavily on access to full-text content or well-annotated citation metadata.

The Data-index \cite{hood2021data} quantifies both dataset production and documented reuse events. It emphasizes reproducibility and rewards data sharing by explicitly tracking reuse, a feature lacking in traditional author-centric metrics. However, it does not account for transitive reuse that is, when datasets are reused indirectly through derived products, nor does it fully capture the influence of datasets reused in interdisciplinary or applied contexts. The SCIENCE-index \cite{adams2023assessing} was introduced to improve transparency. It incorporates blockchain technology to securely record dataset generation, access, and reuse, thus enhancing trust in credit attribution. It combines researcher career statistics with dataset reuse counts to enable a more holistic impact assessment. Despite these innovations, the SCIENCE-index faces two significant limitations: (1) the computational and energy costs of blockchain-based infrastructure, which may not be scalable for large or decentralized scientific ecosystems, and (2) the need for field-specific calibration, as disciplinary norms vary significantly in how data is cited, reused, or acknowledged. The T-index \cite{cerchiello2016tweets} modified the h-index by integrating retweet counts and statistical confidence intervals. This provides a more stable and robust measure of user influence, particularly in real-time and volatile communication networks. While the T-index contributes to measuring engagement and public discourse, its design is highly domain-specific. It has not been generalized to scientific datasets or scholarly contexts, and it inherently depends on the dynamics of social media platforms, which may not reflect scholarly or long-term research impact.

Overall, these approaches highlight three unresolved limitations: (i) a reliance on raw citation counts or surface-level usage statistics, and (ii) the lack of mechanisms to capture dataset accessibility and downstream impact through transitive reuse, and (iii) the absence of a unified metric that can be consistently applied across disciplines. Our proposed X-index addresses these gaps by explicitly integrating breadth, quality, citation impact, and reuse depth into a unified framework for recognizing and rewarding dataset contributions.

\section{Proposed formulation}
In this section, we present our proposed metric—the X-index (Cross-domain index), designed to quantify the impact of data sharing by capturing both dataset-level value and author-level contribution. Our approach addresses key limitations of existing citation-based or reuse-specific metrics by integrating multidimensional indicators of dataset utility, quality, and downstream influence. We follow the FAIR principles (Findability, Accessibility, Interoperability, and Reusability) as a major parameter to score the dataset. Findable refers to the metadata and data being easy to find. Accessibility refers to the conditions under which the data is easily accessible and understandable. Interoperable refers to the standardization of vocabularies, ontologies, thesauri, and so on, such that they can easily be integrated into the existing systems. Reusable refers to the need for metadata and data to be clearly and comprehensively described, ensuring they can be easily replicated, integrated, or combined with existing systems. The proposed X-index is calculated in two steps:
\begin{itemize}
    \item Computation of V-score (Value Score) for each dataset
    \item Aggregation into X-index at the author level
\end{itemize}

For each published dataset, we compute a V-score $V_i$, which quantifies the degree to which the dataset contributes to scientific progress. The V-score incorporates four components, each reflecting a critical dimension of dataset value:
\begin{itemize}
\item {Breadth (X)}: This measures the disciplinary diversity and collaborative reach of a dataset. It captures how many distinct research domains or unique co-authors across fields have reused the dataset, incentivizing cross-disciplinary impact and openness.
\item {Quality (Y)}: Represented by a proxy FAIR score, this component evaluates the dataset’s compliance with the principles of Findability, Accessibility, Interoperability, and Reusability. It is determined based on public availability in recognized repositories, presence of metadata, and adherence to data-sharing standards. 
\item {Citation Impact (C)}: This is the number of citations that either directly cite the dataset or the publication in which the dataset was introduced.
\item {Reuse Depth (D)}: This is a decay-weighted transitive reuse factor, representing the depth of reuse through downstream research. It captures how the dataset propagates influence through chains of citations. Direct citations are weighted at 1, first-order (one-hop) citations at ½, second-order at ¼, and so on, with a decay cap at four levels for computational feasibility.
\end{itemize}

These components are additively combined into the final value score for the $i^{th}$ dataset as follows:
\begin{equation}
    V_i = X_i + Y_i + D_i \cdot \log\!\left(1 + C_i\right)
\end{equation}

This formulation introduces a logarithmic function over citation count $C_i$ to dampen the disproportionate influence of highly cited datasets while still rewarding citation growth. Crucially, the inclusion of $D_i$ acknowledges indirect influence, a major blind spot in traditional citation-based metrics. We calculate the reuse depth multiplier using decay weights (1 for direct citations, ½ for one-hop, ¼ for two-hop, etc.), effectively tracing the dataset’s impact as it propagates through citation chains. 

To assess individual author contributions to open and impactful data sharing, we define the cross domain index (X-index) as the aggregate of value of the datasets the author has published. It is defined as:
\begin{equation}
    X-Index = \sum_i V_i
\end{equation}

Here, the summation runs over all datasets, on which the author is listed as a co-author. The X-index thus reflects not only the volume of datasets shared but also their actual reuse, influence, and quality, providing a comprehensive measure of scholarly openness and data-centric impact.

This formulation contrasts with traditional author metrics that often focus solely on publication counts or direct citations. By emphasizing the characteristics and reuse behaviors of the data set, the X-index provides a contribution-sensitive, reuse-driven, and interdisciplinary-aware evaluation of the scholarly contribution.

\subsection{Computational implementation}
As illustrated in Figure~\ref{fig:pip}, our proposed framework consists of multiple components to collect bibliometric and dataset usage data, which are then used to compute two levels of impact scores: the V-score (dataset-level) and the X-index (author-level). Data were gathered through a combination of publicly accessible sources and APIs, namely Google Scholar \footnote{\url{https://colab.research.google.com/}}, SerpAPI \footnote{\url{https://www.searchapi.io/}}, and OpenAlex \footnote{\url{https://openalex.org/}}.

We developed the entire processing pipeline using Python (version 3.10), utilizing libraries such as numpy, pandas, and so on for data manipulation, numerical operations, and integration with APIs. The pipeline follows a modular structure that allows for scalable and reproducible computation of impact metrics across a wide range of datasets and authors.

The computation of the V-score, a composite metric for dataset-level impact, is based on four key components: breadth, quality, citation impact, and reuse depth. These components are described in detail below:
\begin{figure}[h]
\centering
\includegraphics[height=8cm]{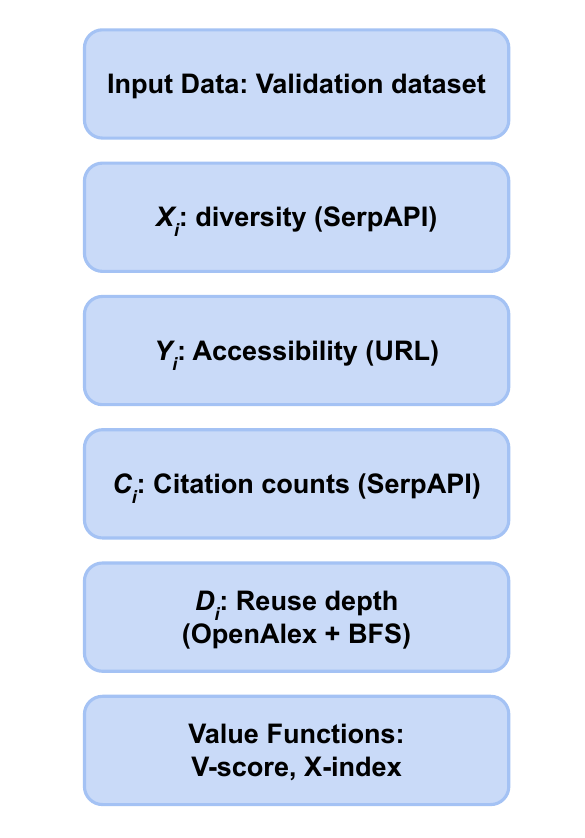}
\caption{An illustration of our V-score and X-index pipeline.}
\label{fig:pip}
\vspace{-10pt}
\end{figure}

\begin{itemize}
\item Breadth ($X_i$): To assess the disciplinary diversity of a dataset's usage, we first extract the list of co-authors associated with publications that cite or use the dataset. This is achieved using the OpenAlex and SerpAPI services. For each dataset, we calculate the normalized Shannon entropy of the disciplinary distribution of these co-authors (based on their affiliations and publication venues), providing a quantitative measure of diversity. In addition to entropy, we also consider the total number of unique fields represented. To avoid instability due to datasets with extremely low diversity, a lower bound (floor) is applied to the entropy values.
\item Quality ($Y_i$): This dimension reflects the accessibility and usability of the dataset. Specifically, we assign quality weights based on the availability of a functional and verifiable URL that provides access to the dataset. A binary weighting scheme is applied—datasets with a valid, publicly accessible URL.

\item {{Citation Impact ($C_i$)}}: Citation counts are computed based on SerpAPI's Google Scholar search. Specifically, counts are scaled by reuse depth and transformed by log to stabilize variance. 
\item {{Reuse Depth ($D_i$)}}: This metric captures the extent to which a dataset has been reused indirectly through citation chains. We construct a citation graph using data from OpenAlex, treating each publication as a node and citation relationships as edges. Starting from a seed publication, we perform a breadth-first search (BFS) over the graph. The reuse depth is quantified by the number of citation “hops” from the seed publication and the number of citing works discovered at each depth level. This allows us to account not only for direct citations but also for second-order and deeper reuses of the dataset.
\end{itemize}

Finally, we summed the V-scores across all the datasets to derive the X-index for each author. 

\subsection{Feasibility}
The X-index is designed with practical implementation in mind, ensuring that it is both feasible and accessible across a broad range of research disciplines. A key strength of the X-index is its low computational cost and the use of widely available technologies and resources. The model is computationally lightweight, enabling its deployment on standard consumer-grade devices without the need for expensive infrastructure. Specifically, the X-index relies on free software tools such as Google Colab and Visual Studio Code\footnote{\url{https://code.visualstudio.com/}}
 within a Python environment. Additionally, it utilizes open-access APIs, such as SerpAPI and OpenAlex, that are freely available to the research community, further reducing the barriers to entry.

The core formula of the X-index is designed to be both simple and efficient, ensuring that it can be implemented without significant computational overhead. This makes it a viable option for researchers and institutions with limited access to high-performance computing resources. Importantly, because the algorithm is lightweight, it is suitable for use on average laptops or desktop computers. This approach ensures that the X-index is cost-effective to implement, making it accessible to a wide range of research institutions, including those in resource-constrained environments.

To assess the computational efficiency and scalability of the X-index, we conducted a preliminary experiment involving 15 datasets from nine distinct authors. The core computation—the V-score took approximately six seconds per dataset. After this, computing the X-index from the V-scores required just under ten additional seconds, bringing the total processing time for all datasets to under 100 seconds. These results were achieved using a low-power machine that utilized only 1.1 GB of RAM out of a total of 12.7 GB, illustrating the X-index's efficiency even on machines with modest capabilities. Given the model's simplicity and the computational efficiency demonstrated in this experiment, it is expected that the X-index will be able to handle much larger datasets (e.g., datasets exceeding 100 entries) without significant increases in processing time or resource consumption.

Additionally, the underlying databases that feed into the X-index evolve slowly over time. As a result, we propose utilizing incremental computation to periodically update the index, allowing for efficient recalculation of the X-index without imposing excessive computational burden. This approach ensures that the X-index remains responsive to changes in the underlying data while minimizing unnecessary computational effort.

These characteristics of the X-index, low resource requirements, computational simplicity, and scalability make it a highly feasible solution for diverse research domains. Furthermore, the ability to implement the X-index on widely available platforms without the need for specialized infrastructure makes it accessible to a broad user base, from individual researchers to large research institutions.

\subsection{Impact}
The X-index is not only feasible but also impactful in reshaping how datasets are evaluated and recognized in academic, policy, and industry contexts. By addressing critical gaps in traditional citation-based metrics, the X-index offers a more comprehensive, equitable, and transparent system for evaluating dataset contributions. Specifically, the X-index integrates both citation frequency and reuse depth, enabling a more nuanced understanding of dataset impact.

Traditional citation counts have long been the gold standard for academic impact assessment. However, these metrics fail to capture the full scope of a dataset's influence. Citation counts only reflect the number of times a dataset is cited, which does not account for how widely and deeply the dataset is reused in subsequent research. In contrast, the X-index goes beyond citation counts by incorporating reuse depth—measuring how extensively and in how many different contexts the dataset is used. This dual focus on both the breadth (the number of projects or researchers using the dataset) and depth (the extent to which the dataset influences subsequent research) allows the X-index to provide a more holistic view of a dataset’s academic impact. For example, datasets that are reused across multiple research domains and by a wide variety of researchers will receive higher V-scores, which reflect not only their citation count but also the importance of their reuse. This broader evaluation framework makes the X-index particularly attractive for researchers who wish to receive fair, data-driven recognition for their contributions to the scientific community, irrespective of their citation history or academic status.

Furthermore, the X-index aligns with the objectives of major funding agencies, which are increasingly promoting open science and the reuse of publicly funded research datasets. By incentivizing dataset reuse and making the impact of datasets more transparent, the X-index supports the principles of open data, which are crucial for fostering collaboration and accelerating scientific progress. These agencies have emphasized the need to ensure that publicly funded datasets are accessible, reusable, and used by the broader scientific community. The X-index incentivizes researchers to make their datasets open, ensuring that they can be freely shared and reused, thereby contributing to the broader goals of open science. By linking dataset excellence to measurable scores, the X-index also encourages researchers to enhance the FAIRness of their datasets and improve their documentation practices. This commitment to data quality enhances the overall scientific ecosystem by promoting the reuse of well-documented and high-quality datasets, thus fostering a culture of proactive data stewardship.

The X-index also addresses significant concerns about the limitations of traditional citation metrics, which can be biased toward well-established researchers or large research teams. By incorporating reuse depth into its evaluation, the X-index mitigates these biases, recognizing the value of datasets from less-visible authors or smaller research groups whose datasets might not be widely cited but are crucial for advancing scientific knowledge. This makes the X-index a more inclusive and equitable metric, ensuring that datasets from all sources are fairly evaluated based on their actual impact.

Moreover, the X-index uses a combination of direct citation tracking and decay-weighted transitive citations, a methodology informed by network-based citation models \cite{walker2007ranking,yan2012scholarly}. This dual approach ensures that both the immediate impact (measured by direct citations) and the long-term, cross-disciplinary impact (measured by the transitive reuse of datasets) are captured and rewarded. Datasets that achieve widespread reuse, especially across different research disciplines, are given higher V-scores, reflecting their foundational importance. The X-index aggregates these high V-scores to recognize researchers who consistently produce datasets with a lasting impact on the research community.

The impact of the X-index is not confined to academic recognition alone. By introducing a model that rewards data sharing and incentivizes dataset reuse, the X-index contributes to the broader push for open science and greater collaboration in the research community. Through these mechanisms, the X-index promotes a more transparent, equitable, and sustainable model of academic credit and incentivizes researchers to contribute to the development of high-quality, reusable datasets. In turn, this fosters a culture of data stewardship, where datasets are treated as valuable research outputs worthy of sustained care, reuse, and community involvement.

\section{Validation of Formula}
To validate the proposed V-score formula both quantitatively and qualitatively, we conducted an initial validation experiment using a diverse sample of datasets from various research domains. The sample comprised 15 datasets authored by 9 different researchers, covering disciplines such as medicine, computational social sciences, and crisis communication. These datasets had a wide range of citation counts, varying from 0 to 8,553, ensuring a broad representation of citation impact. The datasets were evaluated by 5 independent researchers from different domains, who assessed each dataset on four key parameters:
\begin{itemize}
    \item Diversity of the domain: The breadth of topics or disciplines impacted by the dataset.
    \item FAIR principles: The extent to which the dataset adheres to the FAIR principles (Findability, Accessibility, Interoperability, and Reusability).
    \item Citation: The number of citations the dataset has received, which traditionally measures its impact.
    \item Reusability: The degree to which the dataset has been reused across different research contexts, reflecting its broader academic value
\end{itemize}

\begin{figure}[h]
  \centering
  \includegraphics[height=6cm]{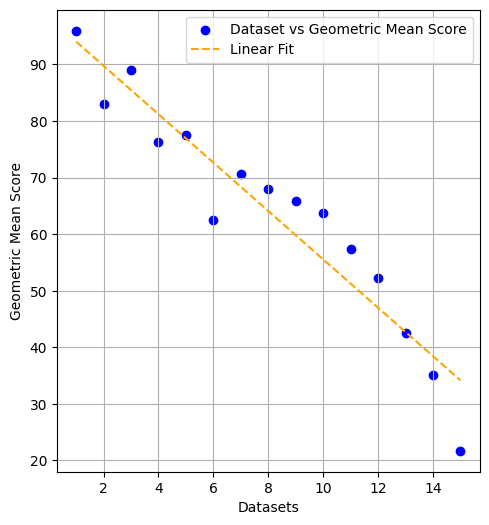}
  \caption{Regression analysis: Geometric Mean Score}
  \label{geometric mean}
\end{figure}

\begin{figure}[h]
  \centering
  \includegraphics[height=6cm]{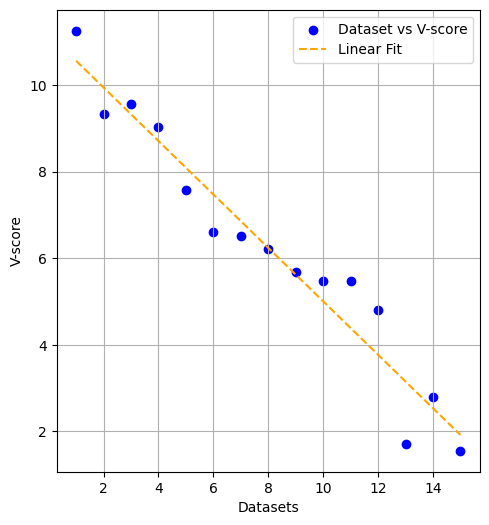}
  \caption{Regression analysis:V-score}
  \label{vscore}
\end{figure}

Each parameter was scored on a scale from 0 to 25, and the individual scores for each dataset were then aggregated on a 0-100 scale. The aggregated score from each rater is referred to as the rater-score. This method allows for a more comprehensive evaluation that includes not only citation counts but also the broader impact and quality aspects of the dataset.

To account for any potential outliers or skewed distributions in the rater scores, we calculated the geometric mean of the scores from all five raters for each dataset. The geometric mean was chosen because it provides a more robust measure when the data is not normally distributed, making it less sensitive to extreme values. The rater-ranking of datasets was then established based on these geometric mean scores, with datasets ranked on a scale from 1 to 15, where 1 represented the highest-rated dataset and 15 the lowest.

These rater-scores and rater-rankings provided our gold standard for comparing the accuracy and validity of the V-score formula.

Next, we computed the V-scores for each dataset using the proposed formula. These V-scores were then used to rank the datasets, and the resulting V-score ranking was compared to the rater-ranking to assess how well the V-scores aligned with the human raters’ evaluations. The comparison was made using Spearman’s rank correlation, which measures the strength and direction of the monotonic relationship between two rankings. We found a strong Spearman correlation of 0.95, indicating that the V-score rankings closely mirrored the rankings provided by human raters, demonstrating high reliability and consistency in our formula.

To further assess the relationship between the V-scores and the gold standard rankings, we performed a regression analysis (Figure~\ref{geometric mean} and Figure~\ref{vscore}). The goal was to determine how well the V-scores predicted the geometric mean of the rater-scores, which serves as a reliable measure of dataset quality. The results of the regression showed a strong linear fit, with an $R^2$ value of 0.94 for the V-scores. This indicates that the V-scores explain 94\% of the variation in the geometric mean, confirming that our formula accurately reflects the human raters’ overall evaluation.

In addition to the $R^2$ value, the slope of the regression line is an important indicator of how the V-scores relate to the rater-scores. The slope of the V-score regression line was calculated to be -0.6171, while the slope of the geometric mean regression was -4.2755. These slopes were rounded to two decimal places, yielding a score relation of approximately 7:1 (geometric mean: V-score). This ratio demonstrates that the V-score values are well-aligned with the human rater scores, confirming that the proposed formula performs robustly in predicting dataset impact based on the raters’ multi-dimensional assessments.

In conclusion, the initial validation results strongly support the validity and reliability of the V-score formula. The high Spearman correlation and the excellent linear fit between V-scores and rater-scores indicate that the proposed formula can effectively capture both the academic impact (via citation count) and the broader value (via reuse and FAIRness) of datasets.

\vspace{-2pt}
\section{Discussion}
The results of this validation experiment demonstrate that the proposed V-score formula is both reliable and valid for assessing dataset impact. The high Spearman correlation and the strong linear fit between V-scores and human evaluations suggest that the formula can accurately rank datasets based on their citation frequency, reuse depth, and overall quality. This is particularly important because traditional citation-based metrics often fail to capture the full scope of a dataset's impact, especially in terms of its reuse in subsequent research. By integrating both citation counts and reuse depth, the V-score formula offers a more comprehensive and holistic approach to dataset evaluation.

The $R^2$ value of 0.94 further supports the robustness of the V-score, indicating that it accounts for a substantial portion of the variation in dataset impact as judged by the human raters. This strong correlation suggests that the V-score is not only a theoretical construct but also a practical and effective tool for evaluating datasets in real-world academic contexts. The 7:1 ratio of geometric mean to V-score slope also provides useful insight into the relative weight of citation counts versus reuse depth in the formula. This ratio indicates that while citation counts are important, the formula gives significant weight to the broader impact of datasets, as measured by their reuse in various research contexts.

One important consideration is that the validation dataset used in this study was relatively small (15 datasets), which may limit the generalizability of these findings. However, the strong correlation and linear regression results suggest that the V-score formula is likely to be effective across a broader range of datasets. In future work, we plan to test the formula on a larger and more diverse set of datasets to confirm that the results hold up at scale. Additionally, we will explore the impact of different weights for the parameters (e.g., citation vs. reuse depth) to further refine the formula and ensure that it is adaptable to various research domains.

The proposed X-index, which aggregates V-scores to rank authors, shows promise as a metric for recognizing dataset contribution at the researcher level. Our current validation focuses on the V-score at the dataset level, but future work will examine how the X-index can be used to rank authors or research groups, providing a more comprehensive evaluation of individual contributions to the scientific community.

Importantly, the transparency and low computational cost of the V-score formula make it an attractive alternative to traditional citation-based metrics. The formula can be easily implemented using widely available tools like Google Colab and Visual Studio Code, making it accessible even for resource-constrained researchers and institutions. This aligns with the growing emphasis on open science and the need for metrics that encourage data sharing and reuse. By rewarding datasets for their impact in both citation and reuse, the V-score provides a fairer and more inclusive way to recognize research contributions, especially for less-visible datasets and emerging research areas that may not yet have high citation counts.

Moreover, the scalability of the V-score formula is another key strength. As demonstrated in our experiment, the V-score formula can be computed efficiently even on low-power machines, making it feasible to handle large datasets without significant computational overhead. This scalability is particularly important as datasets continue to grow in size and complexity, especially in fields like computational social sciences and medicine where data sharing is crucial.

% \vspace{-8pt}
\section{Conclusion}
In this work, we introduced the X-index, a novel metric that measures the value of a researcher’s entire body of datasets over the course of their career. The X-index is calculated through a two-step process: (i) the computation of a Value Score (V-score) for each dataset, based on several factors that reflect its academic impact and reuse potential, and (ii) the aggregation of individual V-scores into the X-index (Cross-domain Index), which represents the cumulative value of all datasets produced by a researcher.

The X-index offers a straightforward, transparent, and computationally efficient approach to evaluating the long-term contributions of researchers in terms of their datasets. In our initial validation, which involved comparing the X-index with a gold standard established by human raters, we observed a strong correlation, demonstrating that the metric reliably captures the value of datasets as judged by experts.

These promising results provide strong evidence of the X-index’s potential as a fair and effective tool for recognizing dataset contributions in academic research. The simplicity and scalability of the formula make it easy to implement, even on modest computational resources, while offering a more comprehensive and nuanced view of a researcher’s impact compared to traditional citation-based metrics.

Given the growing emphasis on open science and the need for better recognition of data-sharing practices, we are optimistic that the X-index could serve as a valuable addition to the landscape of research metrics. Moving forward, we are eager to refine and expand this metric further, particularly by testing it on larger datasets and incorporating additional dimensions of dataset impact. Ultimately, our goal is to promote greater transparency, fairness, and collaboration in the research community by providing a more holistic measure of dataset value.



\begingroup
\raggedright
\sloppy
\bibliographystyle{ACM-Reference-Format}
\bibliography{sample-base}
\endgroup

% --- Compact appendix on same page ---
\clearpage
\appendix
\section{Appendix A}

\begin{table*}[!b]
\centering
\caption{Computational resources available for calculating X-index on the validation dataset.}
\begin{tabular}{l l}
\hline
\textbf{Resource} & \textbf{Value} \\
\hline
RAM & $\sim$12.7 GB \\
Disk Space & $\sim$107 GB \\
Available Time & Up to 100 hours per session \\
Python Backend & Python 3 Google Compute Engine \\
\hline
\end{tabular}
\label{apptaba}
\end{table*}

\begin{table*}[!b]
\centering
\caption{Technologies, tools, and resources.}
\begin{tabularx}{\textwidth}{l l l X}
\toprule
\textbf{Category} & \textbf{Component} & \textbf{Tool / Technology} & \textbf{Description} \\
\midrule
Data Scraping, input \& APIs & Citation count and coauthor metadata & SerpAPI & Scrape Google Scholar for citation count and coauthor metadata \\
 & Author field/domain & OpenAlex API & Retrieve the research domain for each coauthor \\
\midrule
Programming \& Libraries & Programming environment & Python 3.x & Primary language for automation \\
 & Data processing & Pandas & Load, clean, and process dataset samples \\
 & API querying & requests & Handle API requests \\
 & Entropy \& math & math, collections.Counter & Compute V-score and X-index math \\
 & Visualization & matplotlib & Generate plots for V-score and X-index values \\
\midrule
Data Input Resource & Validation dataset & validation\_data.csv & Curated CSV with datasets, paper title, links, author, Google Scholar ID \\
 & Citation source & Scraped via SerpAPI & Used to compute the reuse depth and breadth \\
\midrule
Output \& Visualization & Charts & PNG / SVG via matplotlib & Line/bar plots for X-Index per author \\
 & Export results & pandas & Contains parameters, V-score, and X-Index along with dataset information \\
\bottomrule
\end{tabularx}
\label{apptabb}
\end{table*}

\end{document}